\documentclass[a4paper,twoside,12pt]{article}
\input{obsjkt.sty}

\newcommand{\Msun}{\ensuremath{~{\rm M}_\odot}}                   % Solar mass symbol
\newcommand{\Rsun}{\ensuremath{~{\rm R}_\odot}}                   % Solar radius symbol
\newcommand{\rhosun}{\ensuremath{~\rho_\odot}}                    % Solar density symbol
\newcommand{\Teff}{\ensuremath{T_{\rm eff}}}                      % Effective temperature symbol
\newcommand{\logg}{\ensuremath{\log g}}                           % log(g) symbol
\newcommand{\degr}{\ensuremath{^\circ}}                           % Degree symbol
\renewcommand{\kms}{~km~s$^{-1}$}                                 % km/s symbol
\renewcommand{\cd}{~d$^{-1}$}                                     % Cycles per day symbol
\newcommand{\hip}{\textit{Hipparcos}}
\newcommand{\gaia}{\textit{Gaia}}
\newcommand{\targ}{GK~Dra}
\newcommand{\targfull}{GK~Draconis}

\newcommand{\Msunnom}{\hbox{$\mathcal{M}^{\rm N}_\odot$}}
\newcommand{\Rsunnom}{\hbox{$\mathcal{R}^{\rm N}_\odot$}}
\newcommand{\Lsunnom}{\hbox{$\mathcal{L}^{\rm N}_\odot$}}

\usepackage{rotating}

\begin{document} %%%%%%%%%%%%%%%%%%%%%%%%%%%%%%%%%%%%%%%%%%%%%%%%%%%%%%%%%%%%%%%%%%%%%%%%%%%%%%%%%%%%%%%%%%%%%%%%%%%%%%%%%%%%%%%%%%%%%%%%%%%%%%%%%%%%
%%%%%%%%%%%%%%%%%%%%%%%%%%%%%%%%%%%%%%%%%%%%%%%%%%%%%%%%%%%%%%%%%%%%%%%%%%%%%%%%%%%%%%%%%%%%%%%%%%%%%%%%%%%%%%%%%%%%%%%%%%%%%%%%%%%%%%%%%%%%%%%%%%%%%

\OBSheader{Rediscussion of eclipsing binaries: \targ}{J.\ Southworth}{2024 February}

\OBStitle{Rediscussion of eclipsing binaries. Paper XVI. \\ The $\delta$\,Scuti\,/\,$\gamma$\,Dor hybrid pulsator GK~Draconis}

\OBSauth{John Southworth}

\OBSinstone{Astrophysics Group, Keele University, Staffordshire, ST5 5BG, UK}

%\OBSabstract{Abs abs abs abs abs abs abs abs abs abs abs abs abs abs abs abs abs abs abs abs abs abs abs abs abs abs abs abs abs abs abs abs abs abs abs abs abs abs abs abs abs abs abs abs abs abs abs abs abs abs abs abs abs abs abs abs abs abs abs abs abs abs abs abs abs abs abs abs abs abs abs abs abs abs abs abs abs abs abs abs abs abs abs abs abs abs abs abs abs abs abs abs abs abs abs abs abs abs abs abs abs abs abs abs abs abs abs abs abs abs abs abs abs abs abs abs abs abs abs abs abs abs abs abs abs abs abs abs abs abs abs abs abs abs abs abs abs abs abs abs abs abs abs abs abs abs abs abs abs abs abs abs abs abs abs abs abs abs abs abs abs abs abs abs abs abs abs abs abs abs abs abs abs abs abs abs abs abs abs abs abs abs abs abs abs abs abs abs abs abs abs abs abs abs abs abs abs abs abs abs abs abs abs abs abs abs abs abs abs abs abs abs abs abs abs abs abs abs abs abs abs abs abs abs abs abs abs abs abs abs abs abs abs abs abs abs abs abs abs abs abs abs abs.}

\OBSabstract{\targ\ is a detached eclipsing binary system containing two early-F stars, one evolved, in an orbit with a period of 9.974~d and a small eccentricity. Its eclipsing nature was discovered using \hip\ data, and pulsations were found in follow-up ground-based data. Extensive observations have been obtained using the Transiting Exoplanet Survey Satellite (TESS), and we use these and published spectroscopy to perform a detailed reanalysis of the system. We determine masses of $1.421 \pm 0.012$ and $1.775 \pm 0.028$\Msun, and radii of $1.634 \pm 0.011$ and $2.859 \pm 0.028$\Rsun. The secondary component is more massive, larger, and slightly cooler than its companion; the eclipses are total. The properties of the system can be matched by theoretical predictions for an age of 1.4~Gyr and a slightly sub-solar metallicity. We measure 15 significant pulsation frequencies in the TESS light curve, of which three are in the frequency domain of $\gamma$~Doradus pulsations and the remaining 12 are $\delta$~Scuti pulsations; the system is thus a hybrid pulsator. The strongest pulsation can be definitively assigned to the secondary star as it has been detected in radial velocities of this object. TESS will observe \targ\ again for ten consecutive sectors in the near future.}

%%%%%%%%%%%%%%%%%%%%%%%%%%%%%%%%%%%%%%%%%%%%%%%%%%%%%%%%%%%%%%%%%%%%%%%%%%%%%%%%%%%%%%%%%%%%%%%%%%%%%%%%%%%%%%%%%%%%%%%%%%%%%%%%%%%%%%%%%%%%%%%%%%%%%

\section*{Introduction}

Eclipsing binary star systems contain the only stars for which a direct measurement of their most basic properties (mass and radius) is obtainable. Detached eclipsing binaries (dEBs) are particularly useful because their components have evolved as single stars so can be compared to the predictions of theoretical models of stellar evolution, both to check how well the models perform and to guide their improvement \cite{HiglWeiss17aa,ClaretTorres18apj,Tkachenko+20aa}.

Another approach to constraining the theoretical descriptions of stars is via asteroseismology \cite{Aerts++10book}, which uses the measurement of stellar oscillation frequencies to constrain properties such as their densities, ages and rotational profiles \cite{Aerts+03sci,Briquet+07mn,Garcia+13aa,Bedding+20nat}.

A significant fraction of stars are known to show the signatures of both eclipses and pulsations in their light curves. Many of these identifications are a result of the widespread availability of high-quality light curves from space-based telescopes \cite{Me21univ}. The most common class of pulsations seen in dEBs is the $\delta$\,Scuti type \cite{GaulmeGuzik19aa,Chen+22apjs,Kahraman+23mn}, which are short-period pulsations (0.015 to 0.33\,d \cite{Breger00aspc,Grigahcene+10apj}) with pressure as the restoring force. A smaller number show $\gamma$\,Doradus pulsations \cite{Debosscher+13aa,MeVanreeth22mn}, which have longer periods (0.3\,d to 4\,d \citep{Grigahcene+10apj,Henry++07aj}) and gravity as their restoring force. The $\delta$\,Scuti and $\gamma$\,Dor phenomena can occur simultaneously in late-A and early-F stars, examples of which are labelled as hybrid pulsators \cite{Grigahcene+10apj,Balona++15mn}.

In this work we present an analysis of \targfull\ based on published spectroscopy and new space-based photometry. \targ\ is a dEB known to display $\delta$\,Scuti pulsations. We find that it also shows $\gamma$\,Dor pulsations. For further discussion on the motivation of this series of papers see ref.~\cite{Me20obs}.

%%%%%%%%%%%%%%%%%%%%%%%%%%%%%%%%%%%%%%%%%%%%%%%%%%%%%%%%%%%%%%%%%%%%%%%%%%%%%%%%%%%%%%%%%%%%%%%%%%%%%%%%%%%%%%%%%%%%%%%%%%%%%%%%%%%%%%%%%%%%%%%%%%%%%

\section*{\targfull}

\begin{table}[t]
\caption{\em Basic information on \targfull. \label{tab:info}}
\centering
\begin{tabular}{lll}
{\em Property}                            & {\em Value}                 & {\em Reference}                   \\[3pt]
Right ascension (J2000)                   & 16:45:41.19                 & \cite{Gaia21aa}                   \\
Declination (J2000)                       & +68:15:30.9                 & \cite{Gaia21aa}                   \\
%Bright Star Catalogue                    & HR 7551                     & \cite{HoffleitJaschek91}          \\
Henry Draper designation                  & HD 152028                   & \cite{CannonPickering21anhar}     \\
% \textit{Hipparcos} designation          & HIP 97485                   & \cite{Hipparcos97}                \\
% \textit{Tycho} designation              & TYC 2673-65-1               & \cite{Hog+00aa}                   \\
\textit{Gaia} DR3 designation             & 1648575062872337792         & \cite{Gaia21aa}                   \\
\textit{Gaia} DR3 parallax                & $3.2954 \pm 0.0133$ mas     & \cite{Gaia21aa}                   \\          % d = 303.5 +/- 1.2 pc
TESS\ Input Catalog designation           & TIC 230128667               & \cite{Stassun+19aj}               \\
$B$ magnitude                             & $9.12 \pm 0.02$             & \cite{Hog+00aa}                   \\          % \cite{Henden+15aas} for APASS
$V$ magnitude                             & $8.77 \pm 0.01$             & \cite{Hog+00aa}                   \\          % \cite{Hog+00aa} for Tycho
$J$ magnitude                             & $8.001 \pm 0.023$           & \cite{Cutri+03book}               \\
$H$ magnitude                             & $7.886 \pm 0.021$           & \cite{Cutri+03book}               \\
$K_s$ magnitude                           & $7.864 \pm 0.024$           & \cite{Cutri+03book}               \\
Spectral type                             & F1~V + F2~IV                & This work                         \\[3pt]
\end{tabular}
\end{table}

\targ\ (Table~\ref{tab:info}) is one of the 343 eclipsing binaries discovered using data from the \hip\ satellite \cite{Hipparcos97} and named by Kazarovets et al.\ \cite{Kazarovets+99ibvs}. Dallaporta et al.\ \cite{Dallaporta+02ibvs} presented the first ground-based photometry, finding an orbital period of $P = 9.9742$~d, a modest orbital eccentricity, and pulsations in the secondary component consistent with the $\delta$\,Scuti type. The system has since been included in catalogues of binary systems containing $\delta$\,Scuti components \cite{LiakosNiarchos17mn,Kahraman+17mn}. 

Griffin \& Boffin \cite{GriffinBoffin03obs} (hereafter GB03) published the first radial velocity (RV) study of GK~Dra and V1094~Tau (the latter since analysed in detail by Maxted et al.\ \cite{Maxted+15aa}). For \targ\ they obtained 50 RVs of each star. A large scatter in the RVs of the more massive component was found and attributed to the effects of pulsations. A variation of 0.1178~d period was included in the fit to the spectroscopic orbit of the star to account for the pulsation signature. This variation was treated as a Keplerian orbit for convenience, and the fitted eccentricity and velocity amplitudes were $e = 0.26 \pm 0.06$ and $2.62 \pm 0.17$\kms, respectively. The RVs for the less massive star were also found to show an excess scatter indicative of possible pulsations. GB03 found the spectral types of both components to be significantly earlier than the G0 given in the \textit{Henry Draper Catalogue} \cite{CannonPickering21anhar} based on the $B-V$ colour index and the presence of pulsations in at least one of the stars, preferring F2\,III-IV for the more massive star.

Zwitter et al.\ \cite{Zwitter+03aa} (hereafter ZW03) presented the only full analysis of \targ\ published so far. They based their results on data from the \hip\ satellite plus a set of 35 \'echelle spectra covering 848--874~nm, specifically chosen to simulate the type of data expected from the \gaia\ mission \cite{Gaia16aa}. The masses and radii thus determined were $M_{\rm A} = 1.46 \pm 0.07$\Msun, $M_{\rm B} = 1.81 \pm 0.11$\Msun, $R_{\rm A} = 2.43 \pm 0.04$\Rsun\ and $R_{\rm B} = 2.83 \pm 0.05$\Rsun. These numbers indicate that both components are significantly evolved. The RVs from ZW03 are not of the same quality as those from GB03, and the \hip\ photometry is greatly inferior to that now available from TESS, so a reanalysis of \targ\ is warranted.

%2016A&A...591A.118S       [ D ,1]
%Astronomy and Astrophysics, volume 591A, 118-118 (2016/7-1)
%The PASTEL catalogue: 2016 version.
%SOUBIRAN C., LE CAMPION J.-F., BROUILLET N. and CHEMIN L.
%<CDS Catalogue: B/pastel>
%Teff = 6912 pm 80 (Casagrande 2011A&A...530A.138C)

% ok notprinted/1999IBVS.4659....1K.pdf 
% ok writtenup/2002IBVS.5312....1D.pdf 
% ok writtenup/2003Obs...123..203G.pdf 
% ok writtenup/2003A+A...404..333Z.pdf 

%%%%%%%%%%%%%%%%%%%%%%%%%%%%%%%%%%%%%%%%%%%%%%%%%%%%%%%%%%%%%%%%%%%%%%%%%%%%%%%%%%%%%%%%%%%%%%%%%%%%%%%%%%%%%%%%%%%%%%%%%%%%%%%%%%%%%%%%%%%%%%%%%%%%%

\section*{Photometric observations}

\begin{figure}[t] \centering \includegraphics[width=\textwidth]{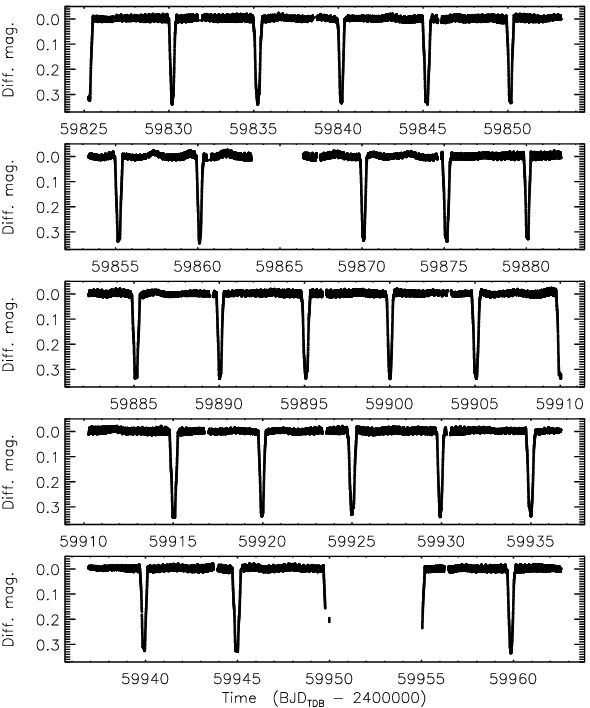} \\
\caption{\label{fig:time} TESS\ short-cadence SAP photometry of \targ. The flux 
measurements have been converted to magnitude units then rectified to zero magnitude 
by subtraction of the median. The individual sectors are labelled.} \end{figure}

\targ\ has been observed extensively by the NASA Transiting Exoplanet Survey Satellite \cite{Ricker+15jatis} (TESS) as it lies in the northern continuous viewing zone of this telescope. Data from sectors 14 to 26 (2019/07/18 to 2020/07/04) were obtained at a cadence of 1800~s, although those from sectors 15 and 16 were not available from the archive at the time of writing. Sectors 40 and 41 (2021/06/08 to 2021/08/20), and 47 to 55 (2021/12/30 to 2022/09/01) yield data at a cadence of 600~s. Finally, in sectors 56 to 60 (2022/09/01 to 2023/01/18) \targ\ was observed at a 200~s cadence.

The data from all the sectors mentioned above were downloaded from the NASA Mikulski Archive for Space Telescopes (MAST\footnote{\texttt{https://mast.stsci.edu/portal/Mashup/Clients/Mast/Portal.html}}) using the {\sc lightkurve} package \cite{Lightkurve18}. The `hard' flag was used to reject data labelled as of lower quality. We used the simple aperture photometry (SAP) data \cite{Jenkins+16spie} for consistency with previous papers in this series. The data were converted to differential magnitude and the median magnitude of each sector was subtracted for convenience.

Our results below are primarily based on the 600-s cadence data as these cover many eclipses whilst avoiding problems with undersampling the light variations of the system. We used the 200-s cadence data for the pulsation analysis due to its higher frequency resolution. The 200-s data are shown in Fig.~\ref{fig:time}, where eclipses and pulsations can both be seen. The 600~s data look similar but of course have a lower sampling rate.

We queried the \gaia\ DR3 database\footnote{\texttt{https://vizier.cds.unistra.fr/viz-bin/VizieR-3?-source=I/355/gaiadr3}} for objects within 2~arcmin of \targ. Only nine were found, and all are fainter by at least 6.3~mag in the $G$ band, so contamination from these objects is negligible.

%%%%%%%%%%%%%%%%%%%%%%%%%%%%%%%%%%%%%%%%%%%%%%%%%%%%%%%%%%%%%%%%%%%%%%%%%%%%%%%%%%%%%%%%%%%%%%%%%%%%%%%%%%%%%%%%%%%%%%%%%%%%%%%%%%%%%%%%%%%%%%%%%%%%%

\section*{Light curve analysis}

\begin{figure}[t] \centering \includegraphics[width=\textwidth]{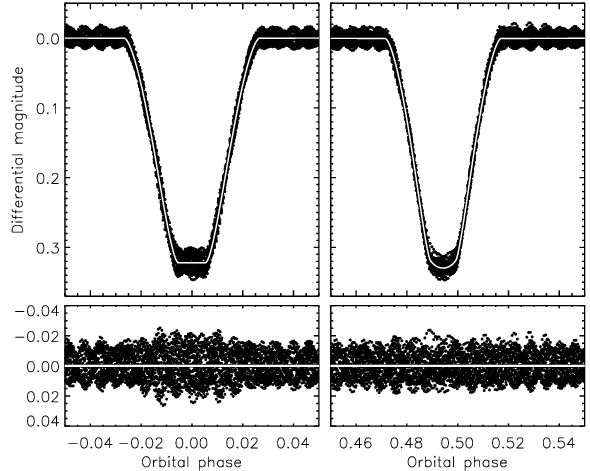} \\
\caption{\label{fig:phase} The 600-s cadence TESS light curves of \targ\ (filled circles) 
and its best fit from {\sc jktebop} (white-on-black line) versus orbital phase. The primary 
eclipse is shown on the left and the secondary eclipse on the right. The residuals are 
shown on an enlarged scale in the lower panel.} \end{figure}

Due to the number and variety of TESS data available for \targ, we investigated two choices with which to develop a model of the system. The first choice was to use only the 200-s cadence data (sectors 56--60) as they have the highest available sampling rate; this was successful but led to a lower precision than desired in the final results. We therefore also modelled the 600-s cadence data from sectors 47--55, augmented by the data from sectors 56--60 reduced to 600~s cadence for consistency.

In both cases we extracted the data within 0.6~d of an eclipse from the full datasets, in order to speed up the computation times in our analysis. The data around each eclipse were individually fitted with a straight line to normalise them to zero differential magnitude, and eclipses which were not fully covered by the data were rejected. This left us with 9730 datapoints (out of 53\,296) for the 200-s light curve and 10\,497 datapoints (out of 47\,378) for the 600-s light curve. In both cases we ignored the errorbars supplied with the TESS data as they do not account for the pulsations. 

We modelled the light curves using version 43 of the {\sc jktebop}\footnote{\texttt{http://www.astro.keele.ac.uk/jkt/codes/jktebop.html}} code \cite{Me++04mn2,Me13aa}. The fitted parameters included the sum ($r_{\rm A}+r_{\rm B}$) and ratio ($k = {r_{\rm B}}/{r_{\rm A}}$) of the fractional radii of the stars ($r_{\rm A}$ and $r_{\rm B}$), the central surface brightness ratio ($J$), orbital inclination ($i$), orbital period ($P$), and a reference time of primary minimum ($T_0$). The orbital eccentricity ($e$) and argument of periastron ($\omega$) were included as $e\cos\omega$ and $e\sin\omega$ to avoid their mutual correlation. Limb darkening was included in our model using the power-2 law \cite{Hestroffer97aa,Me23obs2}: we fitted for the $c$ coefficients and fixed the $\alpha$ coefficients to a theoretically-predicted value \cite{ClaretSouthworth22aa,ClaretSouthworth23aa}. Third light was fixed at zero as attempts to fit it yielded a small negative value. The results are given in Table~\ref{tab:jktebop}.

The lower sampling rate of the 600-s cadence data might bias the measurements of the fitted parameters \cite{Me11mn}. We checked this by running fits where the model was numerically integrated to match the data \cite{Me11mn}, finding a negligible change in the parameter values. We conclude that the sampling rate does not have a significant effect on the results.

The depths of the primary and secondary eclipses are visually inseparable due to the similar \Teff\ values of the stars and the pulsational variability. However, our fits reliably converged to a situation in which the larger and more massive star is slightly cooler, in agreement with the results of ZW03. We therefore define star~A to be the hotter but less massive star -- it is the one eclipsed at the primary (deeper) eclipse. Star~B is thus eclipsed at the secondary eclipse (which occurs at phase 0.4943) and is significantly larger and brighter than star~A. The TESS data reveal that the eclipses are total\footnote{To be precise, the primary eclipse is total and the secondary eclipse is annular.}.

\begin{table} \centering
\caption{\em \label{tab:jktebop} Parameters of \targ\ measured from the TESS light curves using the 
{\sc jktebop} code. The uncertainties are 1$\sigma$ and were determined using residual-permutation 
simulations. We give the results for both the 200-s and 600-s cadence data. We adopted the results 
for the 600-s cadence data but with errorbars double those reported in this table.}
\setlength{\tabcolsep}{4pt}
\begin{tabular}{lcc}
{\em Parameter}                           &      {\em Value (200s)}            &      {\em Value (600s)}             \\[3pt]
{\it Fitted parameters:} \\                                                   
Primary eclipse time (BJD$_{\rm TDB}$)    & $ 2459905.05884  \pm  0.00020    $ & $ 2459905.05884  \pm  0.00024    $  \\
Orbital period (d)                        & $       9.974128 \pm  0.000053   $ & $       9.974128 \pm  0.000012   $  \\
Orbital inclination (\degr)               & $      88.27     \pm  0.11       $ & $      88.467    \pm  0.062      $  \\
Sum of the fractional radii               & $       0.15687  \pm  0.00065    $ & $       0.15641  \pm  0.00042    $  \\
Ratio of the radii                        & $       1.791    \pm  0.023      $ & $       1.750    \pm  0.010      $  \\
Central surface brightness ratio          & $       1.027    \pm  0.048      $ & $       0.938    \pm  0.015      $  \\
% Third light                             & $       0.0001   \pm  0.0008     $ & $       0.0001   \pm  0.0008     $  \\
LD coefficient $c$                        & $       0.670    \pm  0.086      $ & $       0.584    \pm  0.048      $  \\
LD coefficient $\alpha$                   &            0.448 (fixed)           &            0.448 (fixed)            \\
% LD coefficient $c$ for star~A           & $       0.548    \pm  0.017      $ & $       0.548    \pm  0.017      $  \\
% LD coefficient $c$ for star~B           & $       0.516    \pm  0.020      $ & $       0.516    \pm  0.020      $  \\
% LD coefficient $\alpha$ for star~A      &             0.498 (fixed)          &             0.498 (fixed)           \\
% LD coefficient $\alpha$ for star~B      &             0.467 (fixed)          &             0.467 (fixed)           \\
$e\cos\omega$                             & $      -0.008985 \pm  0.000041   $ & $      -0.009000 \pm  0.000036   $  \\
$e\sin\omega$                             & $      -0.07811  \pm  0.0021     $ & $      -0.07998  \pm  0.0017     $  \\
{\it Derived parameters:} \\                                                   
Fractional radius of star~A               & $       0.05620  \pm  0.00028    $ & $       0.05689  \pm  0.00014    $  \\
Fractional radius of star~B               & $       0.10067  \pm  0.00087    $ & $       0.09953  \pm  0.00044    $  \\
Light ratio $\ell_{\rm B}/\ell_{\rm A}$   & $       3.30     \pm  0.24       $ & $       2.87     \pm  0.073      $  \\[3pt]
Orbital eccentricity                      & $       0.0786   \pm  0.0022     $ & $       0.0805   \pm  0.0017     $  \\
Argument of periastron ($^\circ$)         & $     263.44     \pm  0.18       $ & $     263.58     \pm  0.14       $  \\
\end{tabular}
\end{table}

% IDL> print, (0.05689-0.05620)/sqrt(0.00028^2+0.00014^2)
%       2.20412
% IDL> print, (0.10067d0-0.09953d0)/sqrt(0.00087d0^2+0.00044d0^2)
%        1.1693076

For the record, we were able to fit the light curve quite well with the inverse of the ratio of the radii ($k = 0.56$ in this case). This local-minimum solution could be rejected because the limb darkening caused curvature in the wrong eclipse (secondary versus primary) and thus did not match the data.

The uncertainties in the fitted parameters were determined using residual-permutation simulations \cite{Me08mn}, treating the pulsational variation as red noise. Similar uncertainties were found by modelling the  sectors of data individually \cite{Me21obs5}. See our recent paper on V1765~Cyg \cite{Me23obs6} for example plots from a similar residual-permutation analysis. The agreement between the results for the 200-s and 600-s cadence data (Table~\ref{tab:jktebop})is not as good as hoped, with differences between parameters of typically one to two times the size of the uncertainties. As we found above that the lower cadence of the 600-s data was not important, we adopted the results from these data but with the errorbars doubled.

%%%%%%%%%%%%%%%%%%%%%%%%%%%%%%%%%%%%%%%%%%%%%%%%%%%%%%%%%%%%%%%%%%%%%%%%%%%%%%%%%%%%%%%%%%%%%%%%%%%%%%%%%%%%%%%%%%%%%%%%%%%%%%%%%%%%%%%%%%%%%%%%%%%%%

\section*{Pulsation analysis}

\begin{sidewaysfigure} \centering
\includegraphics[width=\textwidth]{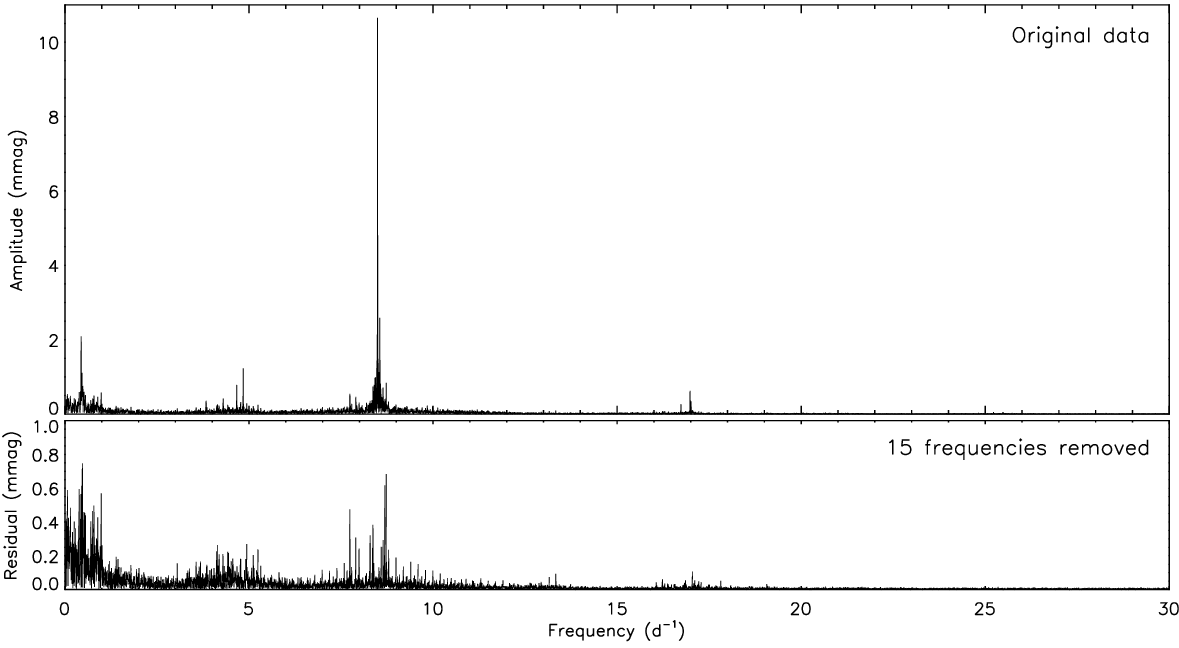}
\caption{\label{fig:freq} Frequency spectrum of the TESS light curve of \targ.
Top: frequency spectrum of the data after subtraction of the binary model.
Bottom: frequency spectrum after subtraction of the binary model and the 15 
frequencies measured in this work.} \end{sidewaysfigure}

The TESS light curve of \targ\ shows clear evidence for pulsations. The strongest frequency ($f_8$) was detected in the RVs of star~B by GB03 so can be unambiguously attributed to this star. We fitted the 200~s cadence data with {\sc jktebop} and subtracted the best-fitting model of the light curve. The residuals of the fit were passed to version 1.2.0 of the {\sc period04} code \cite{LenzBreger05coast} and a frequency spectrum was calculated from 0 to the Nyquist frequency of 216\cd. No significant periodicity was found beyond 30\cd\ (Fig.~\ref{fig:freq}).

We measured a total of 15 significant frequencies in the frequency spectrum, adopting as our significance criterion a signal to noise ratio (SN) greater than 4 (Refs.\ \cite{Breger+93aa},\cite{Kuschnig+97aa}). We then fitted sinusoids simultaneously to all of them to obtain their amplitudes and phases. The uncertainties of the fitted parameters were calculated using both a standard least-square fit and Monte Carlo simulations, the latter being larger. 

% For reference the orbital frequency is 0.1006\cd\ and the Loumos \& Deeming \cite{LoumosDeeming78apss} frequency resolution is $2.5\,/\,\Delta T = 0.095$\cd\ where $\Delta T$ is the time interval covered by the data.

\begin{table} \centering
\caption{\em Significant pulsation frequencies found in the TESS 200-s cadence light
curve of \targ\ after subtraction of the effects of binarity. \label{tab:freq}}
\setlength{\tabcolsep}{12pt}
\begin{tabular}{lccc}
{\em Label} & {\em Frequency (d$^{-1}$)} & {\em Amplitude (mmag)} & {\em Phase} \\[3pt]
$f_{ 1}$ & $ 0.44029 \pm 0.00003$ & $ 2.253 \pm 0.024$ & $0.436 \pm 0.002$ \\ 
$f_{ 2}$ & $ 0.46049 \pm 0.00007$ & $ 1.297 \pm 0.026$ & $0.657 \pm 0.002$ \\ 
$f_{ 3}$ & $ 0.48624 \pm 0.00007$ & $ 0.964 \pm 0.230$ & $0.912 \pm 0.073$ \\ 
$f_{ 4}$ & $ 3.83445 \pm 0.00020$ & $ 0.366 \pm 0.018$ & $0.119 \pm 0.007$ \\ 
$f_{ 5}$ & $ 4.30136 \pm 0.00017$ & $ 0.398 \pm 0.017$ & $0.646 \pm 0.007$ \\ 
$f_{ 6}$ & $ 4.66593 \pm 0.00010$ & $ 0.770 \pm 0.019$ & $0.330 \pm 0.004$ \\ 
$f_{ 7}$ & $ 4.84289 \pm 0.00006$ & $ 1.226 \pm 0.020$ & $0.563 \pm 0.003$ \\ 
$f_{ 8}$ & $ 8.49070 \pm 0.00004$ & $10.574 \pm 0.027$ & $0.135 \pm 0.001$ \\ 
$f_{ 9}$ & $ 8.55349 \pm 0.00003$ & $ 2.191 \pm 0.106$ & $0.578 \pm 0.002$ \\ 
$f_{10}$ & $16.73503 \pm 0.00032$ & $ 0.255 \pm 0.018$ & $0.242 \pm 0.011$ \\ 
$f_{11}$ & $16.98151 \pm 0.00010$ & $ 0.627 \pm 0.289$ & $0.283 \pm 0.004$ \\ 
$f_{12}$ & $17.00896 \pm 0.00020$ & $ 0.322 \pm 0.039$ & $0.788 \pm 0.008$ \\ 
$f_{13}$ & $25.22548 \pm 0.00155$ & $ 0.041 \pm 0.016$ & $0.799 \pm 0.061$ \\ 
$f_{14}$ & $25.47241 \pm 0.00158$ & $ 0.040 \pm 0.019$ & $0.781 \pm 0.063$ \\ 
$f_{15}$ & $25.49961 \pm 0.00150$ & $ 0.042 \pm 0.020$ & $0.137 \pm 0.069$ \\ 
\end{tabular}
\end{table}

The results are given in Table~\ref{tab:freq}. We find three low frequencies ($f_{ 1}$ to $f_{ 3}$) near 0.46\cd\ which are likely of the $\gamma$\,Doradus type. The strongest pulsation ($f_{ 8}$) has a frequency of $8.49070 \pm 0.00004$\cd\ (period $0.1177759 \pm 0.0000006$~d) and amplitude of 10.6~mmag; this is in wonderful agreement with the periodicity of $0.1177753 \pm 0.0000005$~d found in the RVs of star~B by GB03. There are groups of frequencies around 3.8--4.8\cd\ ($f_{ 4}$ to $f_{ 7}$), 8.5\cd\ ($f_{ 8}$ and $f_{ 9}$), 16.7--17.0\cd\ ($f_{10}$ to $f_{12}$) and 25.2--25.5\cd\ ($f_{13}$ to $f_{15}$), of the $\delta$\,Scuti type. The frequency spectum of the residuals in Fig.~\ref{fig:freq} shows excess power in several frequency intervals, suggesting there are additional pulsations below our S/N criterion which might be measurable using additional data.

We conclude that the \targ\ system likely contains at least one hybrid $\delta$\,Sct/$\gamma$\,Dor star -- the caveat here is that we know which is the pulsating star for only one of the frequencies so it is conceivable than one component produces the g-modes and the other the p-modes. Either way, this is an interesting system. None of the frequencies identified here correspond to multiples of the orbital frequency, so we find no evidence for tidally induced or perturbed pulsations. \targ\ has been observed for two sets of 13 consecutive sectors by TESS, with a third one scheduled, so is a good candidate for searching for amplitude modulation in a $\delta$\,Scuti star of known mass and radius \cite{Bowman+16mn}.

We also calculated frequency spectra of the 600-s cadence data from sectors 47 to 60, expecting that the additional data would yield a cleaner spectrum with a lower noise floor. However, the resulting spectra all contained combs of aliases of the strongest frequencies separated by multiples of the orbital frequency; note that these were not see in the 200-s data. This problem occurred using both the residuals of the {\sc jktebop} best fit, the original data, the original data with the eclipses removed, and the original data with the points during eclipse set to zero magnitude. \targ\ will benefit from a more detailed analysis in future, preferably including data from TESS sectors 73 to 83 that are scheduled for observation beginning in December 2023.

%%%%%%%%%%%%%%%%%%%%%%%%%%%%%%%%%%%%%%%%%%%%%%%%%%%%%%%%%%%%%%%%%%%%%%%%%%%%%%%%%%%%%%%%%%%%%%%%%%%%%%%%%%%%%%%%%%%%%%%%%%%%%%%%%%%%%%%%%%%%%%%%%%%%%

\section*{Chromospheric emission}

\begin{figure}[t] \centering \includegraphics[width=\textwidth]{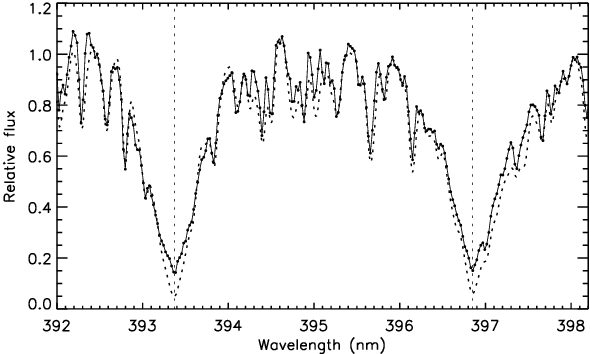} \\
\caption{\label{fig:cahk} Observed spectrum of \targ\ around the Ca~{\sc ii} H and K lines (solid line with points) 
compared to a synthetic spectrum for a star with $\Teff = 7000$~K, $\logg = 4.0$ and solar metallicity from the 
BT-Settl model atmospheres (dashed line) smoothed to the observed spectral resolution. The H and K line central 
wavelengths are shown with dotted lines. Both spectra have been normalised and shifted to zero velocity.} \end{figure}

We obtained a spectrum of the Ca~{\sc ii} H and K lines of \targ, alongside other objects in this series \cite{Me22obs6,Me23obs1}, in order to probe for chromospheric emission lines indicative of magnetic activity. The current target was selected based on its G0 spectral type listed in {\it Simbad}, which is much later than it should be (see above). \targ\ is thus not a promising target for chromospheric emission, but the spectrum only cost about 8~min of observing time. 

The spectrum was obtained on the night of 2022/06/07 in excellent weather conditions, using the Isaac Newton Telescope (INT) and Intermediate Dispersion Spectrograph (IDS), the 235~mm camera, the H2400B grating, the EEV10 CCD, a 1~arcsec slit, and an exposure time of 180~s. It covers 373--438~nm at a reciprocal dispersion of 0.023~nm~px$^{-1}$ and a signal-to-noise ratio of approximately 150, and was taken at an orbital phase of 0.19. Data reduction was performed using a pipeline currently being written by the author \cite{Me+20mn2}.

The spectrum is shown in Fig.~\ref{fig:cahk}, which also includes a synthetic spectrum for a \Teff\ of 7000~K and a \logg\ of 4.0 from the BT-Settl model atmospheres \cite{Allard+01apj,Allard++12rspta}. The Ca H and K line centres have a higher flux than those in the synthetic spectrum, but this can be attributed to the binarity (RV difference at the time of observation) and rotational velocities of the stars. Thus there is no clear evidence for chromospheric emission (as expected).

% IDL> (59900.01455d0-59738.53875d0)/9.974128d0
%        16.189465384843682

%%%%%%%%%%%%%%%%%%%%%%%%%%%%%%%%%%%%%%%%%%%%%%%%%%%%%%%%%%%%%%%%%%%%%%%%%%%%%%%%%%%%%%%%%%%%%%%%%%%%%%%%%%%%%%%%%%%%%%%%%%%%%%%%%%%%%%%%%%%%%%%%%%%%%%

\section*{Physical properties of \targ}

\begin{table} \centering
\caption{\em Physical properties of \targ\ defined using the nominal solar units given by 
IAU 2015 Resolution B3 (ref.\ \cite{Prsa+16aj}). The \Teff\ values are from ZW03. \label{tab:absdim}}
\begin{tabular}{lr@{\,$\pm$\,}lr@{\,$\pm$\,}l}
{\em Parameter}        & \multicolumn{2}{c}{\em Star A} & \multicolumn{2}{c}{\em Star B}    \\[3pt]
Mass ratio   $M_{\rm B}/M_{\rm A}$          & \multicolumn{4}{c}{$1.249 \pm 0.010$}         \\
Semimajor axis of relative orbit (\Rsunnom) & \multicolumn{4}{c}{$28.72 \pm 0.12$}          \\
Mass (\Msunnom)                             &  1.421  & 0.012       &  1.775  & 0.028       \\
Radius (\Rsunnom)                           &  1.634  & 0.011       &  2.859  & 0.028       \\
Surface gravity ($\log$[cgs])               &  4.1642 & 0.0044      &  3.7749 & 0.0083      \\
Density ($\!\!$\rhosun)                     &  0.3257 & 0.0050      &  0.0760 & 0.0020      \\
Synchronous rotational velocity ($\!\!$\kms)&  8.29   & 0.05        & 14.50   & 0.14        \\
Effective temperature (K)                   &  7100   & 70          &  6878   & 57          \\
Luminosity $\log(L/\Lsunnom)$               &   0.786 & 0.018       &  1.217  & 0.017       \\
$M_{\rm bol}$ (mag)                         &   2.774 & 0.045       &  1.698  & 0.042       \\
Distance (pc)                               & \multicolumn{4}{c}{$306.0 \pm 4.8$}           \\[3pt]
\end{tabular}
\end{table}

% IDL> print, [0.012,0.028,0.011,0.028]/[1.421,1.775,1.634,2.859]*100
%      0.844476      1.57746     0.673195     0.979363

Based on the analysis presented above and published results for the system, we have determined the physical properties of \targ. We adopted the values of $r_{\rm A}$, $r_{\rm B}$, $P$ and $i$ from the 600-s cadence data in Table~\ref{tab:jktebop}, doubling the errorbars as described above. For the velocity amplitudes of the system we used the results from GB03 directly as these authors carefully accounted for the effects of pulsations in the RVs: we adopted $K_{\rm A} = 81.14 \pm 0.60$\kms\ and $K_{\rm B} = 64.97 \pm 0.13$\kms\ after interchanging the numbers to account for the differing definitions of which is the primary star. The \Teff\ values were taken directly from ZW03 -- these correspond to spectral types of F1 and F2 on the calibration given by Pecaut \& Mamajek \cite{PecautMamajek13apjs} and are thus much earlier than the G0 given in the \emph{Henry Draper Catalogue} \cite{CannonPickering21anhar}.

The physical properties were then calculated using the {\sc jktabsdim} code \cite{Me++05aa} and entered into Table~\ref{tab:absdim}. The radii are measured to 0.7\% (star~A) and 1.0\% (star~B) precision, limited by the results of the light curve analysis, and the masses to 1.0\% (star~A) and 1.6\% (star~B) precision, limited by the effect of pulsations on the RVs of star~B. The radius measurement of star~A is very different from that of ZW03 (who found $R_{\rm A} = 2.431 \pm 0.042$\Rsun) whereas for star~B the values are consistent (ZW03 obtained $R_{\rm B} = 2.830 \pm 0.054$\Rsun). This can be attributed to the extraordinary improvement in the quality of the TESS light curve versus that previously available, and suggests the older errorbars were significantly underestimated.

% Zwitter:
% R1 = 2.431 ± 0.042
% R2 = 2.830 ± 0.054
% M1 = 1.460 ± 0.066
% M2 = 1.810 ± 0.109

To test our results we determined the distance to \targ\ using the surface brightness calibrations of Kervella et al.\ \cite{Kervella+04aa} for comparison with the \gaia\ parallax. We adopted the apparent magnitudes given in Table\,\ref{tab:info}, but with the 2MASS $JHK_s$ values converted to the Johnson system \cite{Carpenter01aj}. Setting an interstellar reddening of zero gives consistent distances in the five passbands. The most precise is that in $K_s$, $306.9 \pm 4.8$~pc, which we adopt as our final value. This compares favourably with that \gaia\ DR3 \cite{Gaia16aa,Gaia21aa} parallax distance of $303.5 \pm 1.2$~pc, suggesting that the radii and \Teff\ values in Table~\ref{tab:absdim} are reliable. As further evidence, the ratio of the \Teff\ values found by ZW03 is in perfect agreement with the surface brightness ratio we found from the light curve (Table~\ref{tab:jktebop}).

%%%%%%%%%%%%%%%%%%%%%%%%%%%%%%%%%%%%%%%%%%%%%%%%%%%%%%%%%%%%%%%%%%%%%%%%%%%%%%%%%%%%%%%%%%%%%%%%%%%%%%%%%%%%%%%%%%%%%%%%%%%%%%%%%%%%%%%%%%%%%%%%%%%%%%%

\section*{Comparison with theoretical models}

\begin{figure}[t] \centering \includegraphics[width=\textwidth]{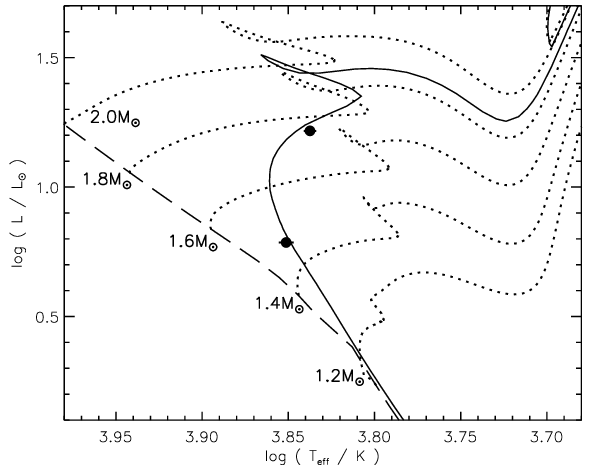} \\
\caption{\label{fig:hrd} Hertzsprung-Russell diagram for the components of \targ\ (filled 
circles with errorbars) and the predictions of the {\sc parsec} 1.2S models for selected 
masses (dotted lines with masses labelled) and the zero-age main sequence (dashed line), 
for a metal abundance of $Z=0.014$. The 1.4~Gyr isochrone is shown with a solid line.} 
\end{figure}

\targ~B shows significant evolution and is now cooler than \targ~A despite its greater mass. We thus decided to compare the measured properties of \targ\ to the predictions of the {\sc parsec} 1.2S theoretical stellar evolutionary models \cite{Bressan+12mn,Chen+14mn}. The best fit was found in the mass--radius and mass--\Teff\ diagrams \cite{MeClausen07aa} for an age of $1400 \pm 50$~Myr and a fractional metal abundance of $Z = 0.014$. The age measurement is very sensitive to the properties of star~B, and once the age is set the \Teff\ of star~A is the primary determinant of the best $Z$ value. The quoted age and $Z$ provide an excellent match to the properties of star~B, but star~A is approximately 2$\sigma$ larger and hotter than predicted. A better agreement could be obtained by interpolation between the $Z=0.010$ and $=0.014$ models, but this is outside the scope of the current work.

We illustrate these results in a Hertzsprung-Russell diagram in Fig.~\ref{fig:hrd}. In this plot are a zero-age main sequence, evolutionary tracks for masses 1.2, 1.4, 1.6, 1.8 and 2.0\Msun, and an isochrone for age 1400~Myr, all for a metal abundance of $Z=0.014$. The isochrone provides a good but not perfect match to the properties of \targ.

%%%%%%%%%%%%%%%%%%%%%%%%%%%%%%%%%%%%%%%%%%%%%%%%%%%%%%%%%%%%%%%%%%%%%%%%%%%%%%%%%%%%%%%%%%%%%%%%%%%%%%%%%%%%%%%%%%%%%%%%%%%%%%%%%%%%%%%%%%%%%%%%%%%%%%%

\section*{Summary and conclusions}

The eclipsing nature of \targ\ was discovered from \hip\ photometry, and subsequent ground-based photometry and spectroscopy allowed the discovery of pulsations in the more massive star and approximate physical properties of this star and its companion. We have revisited \targ\ and used the extensive photometry available from the TESS mission to improve our understanding of the system. We find that the eclipses are total, and determine the masses and radii of the stars to high precision. These match the predictions of theoretical models for an age of 1.4~Gyr and a slightly subsolar metallicity.

The TESS data allow the detection and measurement of 15 significant pulsation frequencies, three of which are have low frequencies in the region of 0.5\cd\ so arise from the $\gamma$\,Dor phenomenon, and the remaining 12 of which form four groups of higher frequencies consistent with $\delta$\,Scuti pulsations. By far the strongest frequency is at $f_8 = 8.49$\cd, and this one has been detected spectroscopically in the more evolved star~B. \targ\ almost certainly contains one or two hybrid $\gamma$\,Dor/$\delta$\,Sct stars, and more extensive data are expected to reveal further pulsation frequencies.

The measured properties of \targ\ are now sufficiently precise and accurate to be included in the Detached Eclipsing Binary Catalogue (DEBCat\footnote{\texttt{https://www.astro.keele.ac.uk/jkt/debcat/}}, ref.~\cite{Me15aspc}). TESS is scheduled to observe \targ\ between December 2023 and September 2024. A detailed asteroseismic analysis of these data, once available, is recommended.

%%%%%%%%%%%%%%%%%%%%%%%%%%%%%%%%%%%%%%%%%%%%%%%%%%%%%%%%%%%%%%%%%%%%%%%%%%%%%%%%%%%%%%%%%%%%%%%%%%%%%%%%%%%%%%%%%%%%%%%%%%%%%%%%%%%%%%%%%%%%%%%%%%%%%%%

\section*{Acknowledgements}

We thank the anonymous referee for a positive report. % quick and 
This paper includes data collected by the TESS\ mission and obtained from the MAST data archive at the Space Telescope Science Institute (STScI). Funding for the TESS\ mission is provided by the NASA's Science Mission Directorate. STScI is operated by the Association of Universities for Research in Astronomy, Inc., under NASA contract NAS 5–26555.
This work has made use of data from the European Space Agency (ESA) mission {\it Gaia}\footnote{\texttt{https://www.cosmos.esa.int/gaia}}, processed by the {\it Gaia} Data Processing and Analysis Consortium (DPAC\footnote{\texttt{https://www.cosmos.esa.int/web/gaia/dpac/consortium}}). Funding for the DPAC has been provided by national institutions, in particular the institutions participating in the {\it Gaia} Multilateral Agreement.
The following resources were used in the course of this work: the NASA Astrophysics Data System; the SIMBAD database operated at CDS, Strasbourg, France; and the ar$\chi$iv scientific paper preprint service operated by Cornell University.

%%%%%%%%%%%%%%%%%%%%%%%%%%%%%%%%%%%%%%%%%%%%%%%%%%%%%%%%%%%%%%%%%%%%%%%%%%%%%%%%%%%%%%%%%%%%%%%%%%%%%%%%%%%%%%%%%%%%%%%%%%%%%%%%%%%%%%%%%%%%%%%%%%%%%

% \bibliographystyle{obsmaga}
% \bibliography{jkt}

\begin{thebibliography}{10}
\newcommand{\enquote}[1]{`(#1)'}

\bibitem{HiglWeiss17aa}
J.~{Higl} \& A.~{Weiss}, \textit{A\&A}, \textbf{608}, A62, 2017.

\bibitem{ClaretTorres18apj}
A.~{Claret} \& G.~{Torres}, \textit{ApJ}, \textbf{859}, 100, 2018.

\bibitem{Tkachenko+20aa}
A.~{Tkachenko} \textit{et~al.}, \textit{A\&A}, \textbf{637}, A60, 2020.

\bibitem{Aerts++10book}
C.~{Aerts}, J.~{Christensen-Dalsgaard} \& D.~W. {Kurtz},
  \textit{{Asteroseismology}} ({Astron.\ and Astroph.\ Library, Springer
  Netherlands, Amsterdam}), 2010.

\bibitem{Aerts+03sci}
C.~{Aerts} \textit{et~al.}, \textit{Science}, \textbf{300}, 1926, 2003.

\bibitem{Briquet+07mn}
M.~{Briquet} \textit{et~al.}, \textit{MNRAS}, \textbf{381}, 1482, 2007.

\bibitem{Garcia+13aa}
A.~{Garc{\'{\i}}a Hern{\'a}ndez} \textit{et~al.}, \textit{A\&A}, \textbf{559},
  A63, 2013.

\bibitem{Bedding+20nat}
T.~R. {Bedding} \textit{et~al.}, \textit{Nature}, \textbf{581}, 147, 2020.

\bibitem{Me21univ}
J.~{Southworth}, \textit{Universe}, \textbf{7}, 369, 2021.

\bibitem{GaulmeGuzik19aa}
P.~{Gaulme} \& J.~A. {Guzik}, \textit{A\&A}, \textbf{630}, A106, 2019.

\bibitem{Chen+22apjs}
X.~{Chen} \textit{et~al.}, \textit{ApJS}, \textbf{263}, 34, 2022.

\bibitem{Kahraman+23mn}
F.~{Kahraman Ali{\c{c}}avu{\c{s}}} \textit{et~al.}, \textit{MNRAS},
  \textbf{524}, 619, 2023.

\bibitem{Breger00aspc}
M.~{Breger}, in \textit{Delta Scuti and Related Stars} ({M.~Breger \&
  M.~Montgomery}, ed.), 2000, \textit{Astronomical Society of the Pacific
  Conference Series}, vol. 210, pp. 3--42.

\bibitem{Grigahcene+10apj}
A.~{Grigahc{\`e}ne} \textit{et~al.}, \textit{ApJ}, \textbf{713}, L192, 2010.

\bibitem{Debosscher+13aa}
J.~{Debosscher} \textit{et~al.}, \textit{A\&A}, \textbf{556}, A56, 2013.

\bibitem{MeVanreeth22mn}
J.~{Southworth} \& T.~{Van Reeth}, \textit{MNRAS}, \textbf{515}, 2755, 2022.

\bibitem{Henry++07aj}
G.~W. {Henry}, F.~C. {Fekel} \& S.~M. {Henry}, \textit{AJ}, \textbf{133}, 1421,
  2007.

\bibitem{Balona++15mn}
L.~A. {Balona}, J.~{Daszy{\'n}ska-Daszkiewicz} \& A.~A. {Pamyatnykh},
  \textit{MNRAS}, \textbf{452}, 3073, 2015.

\bibitem{Me20obs}
J.~{Southworth}, \textit{The Observatory}, \textbf{140}, 247, 2020.

\bibitem{Gaia21aa}
{Gaia Collaboration}, \textit{A\&A}, \textbf{649}, A1, 2021.

\bibitem{CannonPickering21anhar}
A.~J. {Cannon} \& E.~C. {Pickering}, \textit{Annals of Harvard College
  Observatory}, \textbf{96}, 1, 1921.

\bibitem{Stassun+19aj}
K.~G. {Stassun} \textit{et~al.}, \textit{AJ}, \textbf{158}, 138, 2019.

\bibitem{Hog+00aa}
E.~{H{\o}g} \textit{et~al.}, \textit{A\&A}, \textbf{355}, L27, 2000.

\bibitem{Cutri+03book}
R.~M. {Cutri} \textit{et~al.}, \textit{{2MASS All Sky Catalogue of Point
  Sources}} (The IRSA 2MASS All-Sky Point Source Catalogue, NASA/IPAC Infrared
  Science Archive, Caltech, US), 2003.

\bibitem{Hipparcos97}
{ESA} (ed.), \textit{{The Hipparcos and Tycho catalogues. Astrometric and
  photometric star catalogues derived from the ESA Hipparcos space astrometry
  mission}}, \textit{ESA Special Publication}, vol. 1200, 1997.

\bibitem{Kazarovets+99ibvs}
E.~V. {Kazarovets} \textit{et~al.}, \textit{IBVS}, \textbf{4659}, 1, 1999.

\bibitem{Dallaporta+02ibvs}
S.~{Dallaporta} \textit{et~al.}, \textit{IBVS}, \textbf{5312}, 1, 2002.

\bibitem{LiakosNiarchos17mn}
A.~{Liakos} \& P.~{Niarchos}, \textit{MNRAS}, \textbf{465}, 1181, 2017.

\bibitem{Kahraman+17mn}
F.~{Kahraman Ali{\c{c}}avu{\c{s}}} \textit{et~al.}, \textit{MNRAS},
  \textbf{470}, 915, 2017.

\bibitem{GriffinBoffin03obs}
R.~F. {Griffin} \& H.~M.~J. {Boffin}, \textit{The Observatory}, \textbf{123},
  203, 2003.

\bibitem{Maxted+15aa}
P.~F.~L. {Maxted} \textit{et~al.}, \textit{A\&A}, \textbf{578}, A25, 2015.

\bibitem{Zwitter+03aa}
T.~{Zwitter} \textit{et~al.}, \textit{A\&A}, \textbf{404}, 333, 2003.

\bibitem{Gaia16aa}
{Gaia Collaboration}, \textit{A\&A}, \textbf{595}, A1, 2016.

\bibitem{Ricker+15jatis}
G.~R. {Ricker} \textit{et~al.}, \textit{Journal of Astronomical Telescopes,
  Instruments, and Systems}, \textbf{1}, 014003, 2015.

\bibitem{Lightkurve18}
{Lightkurve Collaboration}, \enquote{{Lightkurve: Kepler and TESS time series
  analysis in Python}}, Astrophysics Source Code Library, 2018.

\bibitem{Jenkins+16spie}
J.~M. {Jenkins} \textit{et~al.}, in \textit{Proc.\ SPIE}, 2016, \textit{Society
  of Photo-Optical Instrumentation Engineers (SPIE) Conference Series}, vol.
  9913, p. 99133E.

\bibitem{Me++04mn2}
J.~{Southworth}, P.~F.~L. {Maxted} \& B.~{Smalley}, \textit{MNRAS},
  \textbf{351}, 1277, 2004.

\bibitem{Me13aa}
J.~{Southworth}, \textit{A\&A}, \textbf{557}, A119, 2013.

\bibitem{Hestroffer97aa}
D.~{Hestroffer}, \textit{A\&A}, \textbf{327}, 199, 1997.

\bibitem{Me23obs2}
J.~{Southworth}, \textit{The Observatory}, \textbf{143}, 71, 2023.

\bibitem{ClaretSouthworth22aa}
A.~{Claret} \& J.~{Southworth}, \textit{A\&A}, \textbf{664}, A128, 2022.

\bibitem{ClaretSouthworth23aa}
A.~{Claret} \& J.~{Southworth}, \textit{A\&A}, \textbf{674}, A63, 2023.

\bibitem{Me11mn}
J.~{Southworth}, \textit{MNRAS}, \textbf{417}, 2166, 2011.

\bibitem{Me08mn}
J.~{Southworth}, \textit{MNRAS}, \textbf{386}, 1644, 2008.

\bibitem{Me21obs5}
J.~{Southworth}, \textit{The Observatory}, \textbf{141}, 234, 2021.

\bibitem{Me23obs5}
J.~{Southworth}, \textit{The Observatory}, \textbf{143}, 254, 2023.

\bibitem{LenzBreger05coast}
P.~{Lenz} \& M.~{Breger}, \textit{Communications in Asteroseismology},
  \textbf{146}, 53, 2005.

\bibitem{Breger+93aa}
M.~{Breger} \textit{et~al.}, \textit{A\&A}, \textbf{271}, 482, 1993.

\bibitem{Kuschnig+97aa}
R.~{Kuschnig} \textit{et~al.}, \textit{A\&A}, \textbf{328}, 544, 1997.

\bibitem{Bowman+16mn}
D.~M. {Bowman} \textit{et~al.}, \textit{MNRAS}, \textbf{460}, 1970, 2016.

\bibitem{Me22obs6}
J.~{Southworth}, \textit{The Observatory}, \textbf{142}, 267, 2022.

\bibitem{Me23obs1}
J.~{Southworth}, \textit{The Observatory}, \textbf{143}, 19, 2023.

\bibitem{Me+20mn2}
J.~{Southworth} \textit{et~al.}, \textit{MNRAS}, \textbf{497}, 4416, 2020.

\bibitem{Allard+01apj}
F.~{Allard} \textit{et~al.}, \textit{ApJ}, \textbf{556}, 357, 2001.

\bibitem{Allard++12rspta}
F.~{Allard}, D.~{Homeier} \& B.~{Freytag}, \textit{Philosophical Transactions
  of the Royal Society of London Series A}, \textbf{370}, 2765, 2012.

\bibitem{Prsa+16aj}
A.~{Pr{\v s}a} \textit{et~al.}, \textit{AJ}, \textbf{152}, 41, 2016.

\bibitem{PecautMamajek13apjs}
M.~J. {Pecaut} \& E.~E. {Mamajek}, \textit{ApJS}, \textbf{208}, 9, 2013.

\bibitem{Me++05aa}
J.~{Southworth}, P.~F.~L. {Maxted} \& B.~{Smalley}, \textit{A\&A},
  \textbf{429}, 645, 2005.

\bibitem{Kervella+04aa}
P.~{Kervella} \textit{et~al.}, \textit{A\&A}, \textbf{426}, 297, 2004.

\bibitem{Carpenter01aj}
J.~M. {Carpenter}, \textit{AJ}, \textbf{121}, 2851, 2001.

\bibitem{Bressan+12mn}
A.~{Bressan} \textit{et~al.}, \textit{MNRAS}, \textbf{427}, 127, 2012.

\bibitem{Chen+14mn}
Y.~{Chen} \textit{et~al.}, \textit{MNRAS}, \textbf{444}, 2525, 2014.

\bibitem{MeClausen07aa}
J.~{Southworth} \& J.~V. {Clausen}, \textit{A\&A}, \textbf{461}, 1077, 2007.

\bibitem{Me15aspc}
J.~{Southworth}, in \textit{Living Together: Planets, Host Stars and Binaries}
  (S.~M. {Rucinski}, G.~{Torres} \& M.~{Zejda}, eds.), 2015,
  \textit{Astronomical Society of the Pacific Conference Series}, vol. 496, p.
  321.

\end{thebibliography}

%%%%%%%%%%%%%%%%%%%%%%%%%%%%%%%%%%%%%%%%%%%%%%%%%%%%%%%%%%%%%%%%%%%%%%%%%%%%%%%%%%%%%%%%%%%%%%%%%%%%%%%%%%%%%%%%%%%%%%%%%%%%%%%%%%%%%%%%%%%%%%%%%%%%%
\end{document}